\documentclass[fleqn,12pt,twoside]{article}
\usepackage[headings]{espcrc1}

\readRCS
$Id: espcrc1.tex,v 1.2 2004/02/24 11:22:11 spepping Exp $
\usepackage{graphicx}
\usepackage{subfigure}
\newcommand{\AmS}{{\protect\the\textfont2
  A\kern-.1667em\lower.5ex\hbox{M}\kern-.125emS}}

\title{Hard Probes with the STAR Experiment}

\author{J.C. Dunlop\address[BNL]{Physics Department, \\
        Brookhaven National Laboratory, \\ 
        Upton, NY 11973 U.S.A}%
        (for the STAR\thanks{For the full list of STAR authors and acknowledgements, see appendix `Collaborations' of this volume} Collaboration)}
       
\begin{document}

% typeset front matter
\maketitle

\begin{abstract}
Recent results on the use of hard probes in heavy ion collisions
by the STAR experiment at RHIC are reviewed.  The increased statistical
reach from RHIC run 4 and 
utilization of the full capabilities of the STAR experiment
have led to a qualitative improvement in these results.
Light hadrons have been identified out to transverse momenta ($p_T$) 
of 12 GeV/c, allowing for clear identification 
of the dominant processes governing
particle production in different $p_T$ windows.  
Clean signatures of dijets have been seen even in central Au+Au collisions.
Nuclear modification factors for non-photonic electrons, predominantly
from the decay of heavy-flavored hadrons, have also been measured out
to $p_T$ of 8 GeV/c.   For $p_T > \sim 6$ GeV/c, inclusive spectra 
of all charged hadrons, including heavy-flavored ones, appear
to be suppressed equally strongly (by a factor of four to five)  
in central Au+Au collisions relative
to p+p collisions; interestingly enough, the probability 
of finding a hadron from a dijet partner is suppressed 
to this same level.
\end{abstract}

\section{Introduction}

The use of hard probes in heavy ion collisions has shown 
itself to be highly successful in recent years.  At mid-rapidity,
high transverse momentum ($p_T$) particles are strongly suppressed, by a factor
of four to five, in central Au+Au collisions relative to p+p collisions,
and are not suppressed in 
d+Au collisions~\cite{Adler:2002xw,Adams:2003kv,Adams:2003im}.  
Away-side hadrons from partner dijets are also 
strongly suppressed in central Au+Au collisions,
but not in d+Au collisions~\cite{Adams:2003im,Adler:2002tq}.
The standard explanation is that these phenomena are due to
induced gluon radiation in a dense medium~\cite{GLV,WW,BDMS,Wied}.
Within such approaches, the density needed to reproduce the
data is large, about a factor of
50 beyond that of cold nuclear matter.  

However, significant uncertainties
remain in the determination of the density of the medium through
these methods.  For example, in 
the ``quenching weights'' framework, folded with a realistic geometrical
picture of the overlap zone, the observable single-particle
suppression shows a saturation with increasing density: beyond
a certain density the suppression of light hadrons 
changes rather slowly with increasing
density~\cite{Dainese:2004te,Eskola:2004cr}.  The observed suppression
is deep within this saturated regime, which essentially limits such
measurements to providing a lower bound on the density of the
medium.  More recent questions have arisen as to the contribution
of {\it collisional} rather than radiative energy loss, especially
to the energy loss of heavy quarks~\cite{vanHees:2004gq,Moore:2004tg,Mustafa:2004dr}.
Other related questions are discussed in STAR's recent critical 
assessment~\cite{Adams:2005dq}.  
In order to decrease such uncertainties, more incisive experimental
data is necessary.  

STAR has begun a program to meet this challenge.  First, the experiment
is uniquely positioned to analyze correlations induced by dijets,
due to its full azimuthal coverage.  The measurement
of dihadrons introduces different biases, both geometrical
and fragmentation-induced, than that of inclusive spectra.
Second, with the addition of the Time-of-Flight system (TOF) and Barrel
Electromagnetic Calorimeter (BEMC), STAR is well-positioned to vary the
coupling strength between the probe and the medium, 
which determines the strength of the geometric biases. 
The most dramatic such variation is to set the coupling to zero 
for the trigger particle in a dihadron measurement:
this is the promise of photon-tagged
correlations~\cite{Wang:1996yh}.  An intermediate variation is expected to be
provided by charm and bottom quarks; due to their mass, it has
been predicted that the radiative energy loss of such 
quarks is decreased in medium relative to that of light 
quarks~\cite{Dokshitzer:2001zm,Armesto:2005iq,Djordjevic:2005db}.
The combination of all these different biases 
leads to different dependences
of the suppression on density for the different measurements,
and so it is hoped that the combination of the measurements
will provide a precise measurement of the
density of the medium.

\section{Datasets}
There have been five RHIC runs to date.  Run 4, in 2004, focused
on high statistics measurements in Au+Au at the top collision 
energy, $\sqrt{s_{NN}} = 200$ GeV, along with a smaller set of measurements
at $\sqrt{s_{NN}}$ of 62.4, and a short polarized proton
run at $\sqrt{s} = 200$ GeV.  Roughly an order of magnitude 
more events from $\sqrt{s} = 200$ GeV Au+Au collisions were collected 
in run 4 than in the earlier run 2.  Run 5, in 2005, focused on measurements
in the smaller Cu+Cu system at $\sqrt{s_{NN}}$ of 200, 62.4, and 22 GeV,
along with a longer run with polarized protons at $\sqrt{s} = $ 200 GeV.
Approximately half of the full-energy Au+Au data from run 4 
has been fully reconstructed, along with approximately one-fifth
of the full-energy Cu+Cu data from run 5.  Preliminary results from
these datasets were first presented at this conference.

The STAR experiment has been described in detail elsewhere~\cite{star:nim}.
In runs 4 and 5, significant upgrades were in place.
A large fraction of the BEMC was fully commissioned and active, covering the 
full azimuth for pseudorapidity ($\eta$) $0<\eta<1$.
The BEMC, when combined with the main tracking detector, the Time 
Projection Chamber (TPC), 
enables the measurement of electrons and photons
out to high transverse momentum.
As in run 3, in which the ions collided were d+Au, 
a trigger on high energy in a given calorimeter tower (high tower)
was active, greatly increasing the capability of the
detector to sample the RHIC luminosity and so increasing
the $p_T$ reach of the measurements.  Also, as in run 3,
a small TOF patch, a prototype of the upcoming upgrade to a
full barrel TOF, was in place.  When combined  
with the specific ionization (dE/dx) in the TPC, the TOF 
patch enables measurements of electrons at 
intermediate $p_T$~\cite{Adams:2004fc} and identification
of pions and protons from low $p_T$ ($<\sim 0.5$ GeV/c) 
out to $p_T>\sim$ 12 GeV/c~\cite{Shao:2005iu}.   
 
\section{Cu+Cu}
\begin{figure}
\centering
\includegraphics[width=0.6\textwidth]{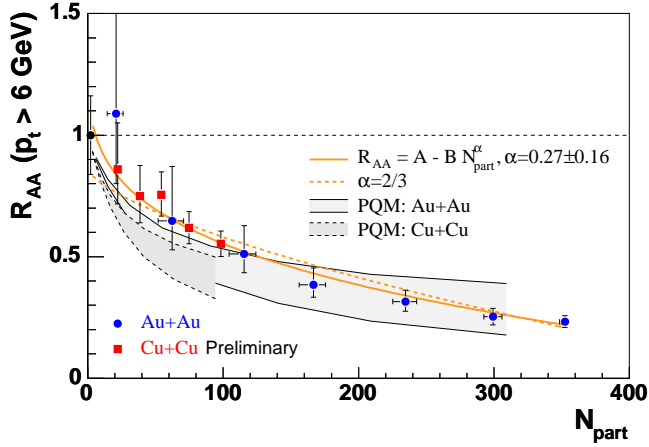}
\caption{Nuclear modification factor $R_{AA}$ for charged hadrons as a function of $N_{part}$.
Au+Au results from ref.~\protect\cite{Adams:2003kv} are shown in blue, while Cu+Cu results are
shown in red.  Uncertainties shown are statistical and point-to-point systematic.
Common uncertainties due to the p+p dataset are shown on the p+p point at left.  PQM predictions are
based on ref.~\protect\cite{Dainese:2004te}.}
\label{fig:cucu_raa}
\end{figure}

As one test of the radiative energy loss picture, in 
run 5 RHIC ran collisions of the lighter system Cu+Cu.
A lighter system brings the advantage that the nuclear
overlap integral $T_{AB}$ is more precisely determined
at low $N_{part}$ than in a heavier system.
Fig.~\ref{fig:cucu_raa} shows the results for the nuclear modification factor $R_{AA}$,
defined as $(dN/dp_T)_{AA}/(T_{AB} (d\sigma/dp_T)_{pp})$, for $p_T>6$ GeV/c, 
as a function of $N_{part}$.  Uncertainties for Au+Au and Cu+Cu datapoints
are dominated by uncertainties in Glauber calculations of $T_{AB}$\cite{Adams:2003yh},
while the common uncertainties from the p+p dataset~\cite{Adams:2003kv} are
placed on the p+p point on the left.  The data show
a clear and common evolution with increasing $N_{part}$: 
for a given $N_{part}$, $R_{AA}$ is
equivalent across systems, though with higher precision in Cu+Cu.
Also shown are phenomenological fits to characterize the $N_{part}$ dependence
of $R_{AA}$.
The data prefer a reduction with the power of $N_{part}^{1/3}$,
though the more commonly expected $N_{part}^{2/3}$ 
reduction~\cite{Vitev:2002pf}
is not strongly excluded.  
This scaling behaviour is a result of a complicated
convolution of the spectral shape and collision geometry. 
The grey bands indicate the results of a full
calculation incorporating such effects~\cite{Dainese:2004te}, which 
reproduces the common suppression
in Cu+Cu and Au+Au at the same $N_{part}$ but gives slightly
larger suppression at low $N_{part}$ than observed in the data.

\section{Baryon enhancement at intermediate $p_T$}
\begin{figure}
\centering
\subfigure[] {
  \label{fig:protonpi}
  \includegraphics[width=0.47\textwidth]{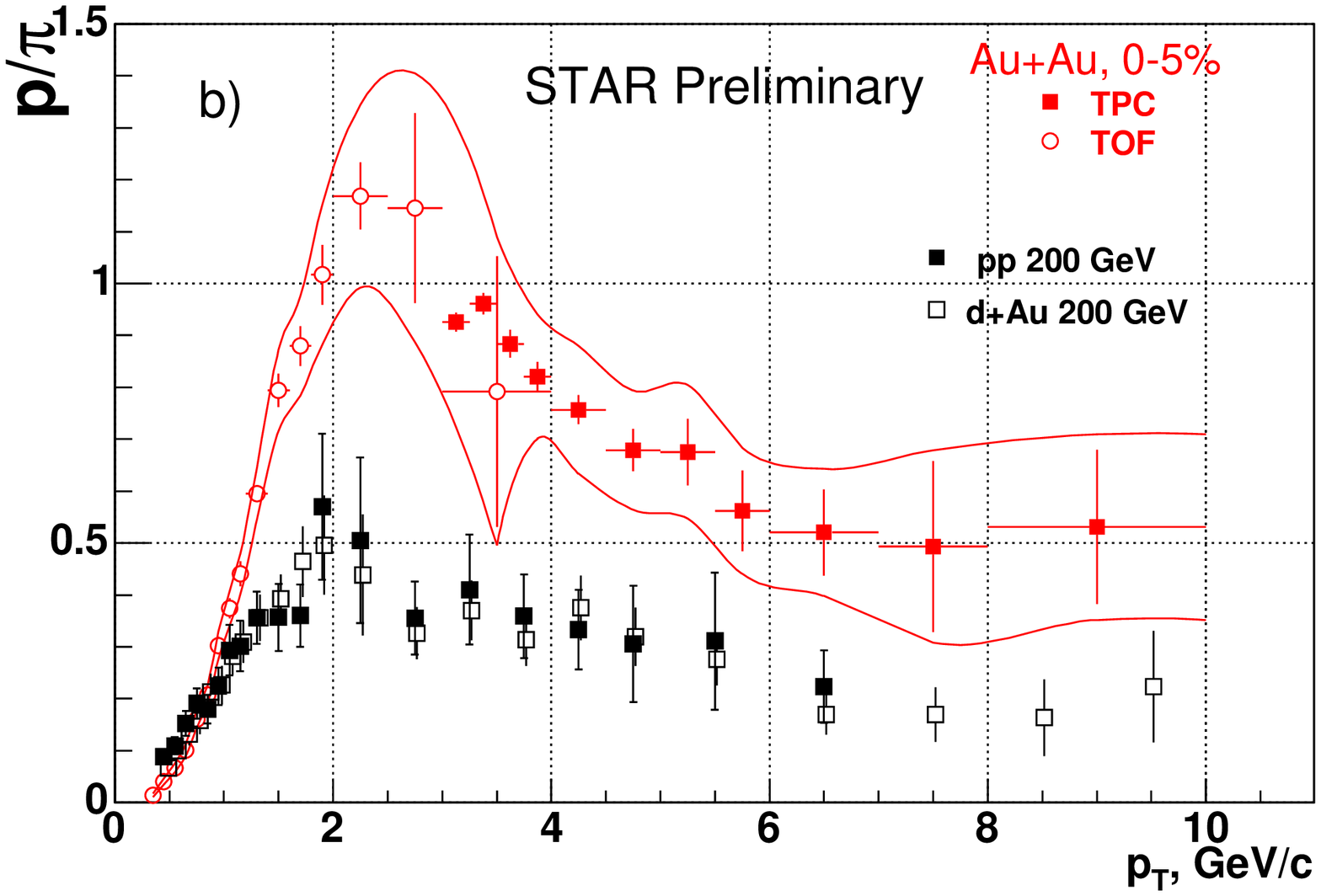}
 }
\hfill
 \subfigure[]
 {
   \label{fig:rcp}
   \includegraphics[width=0.47\textwidth]{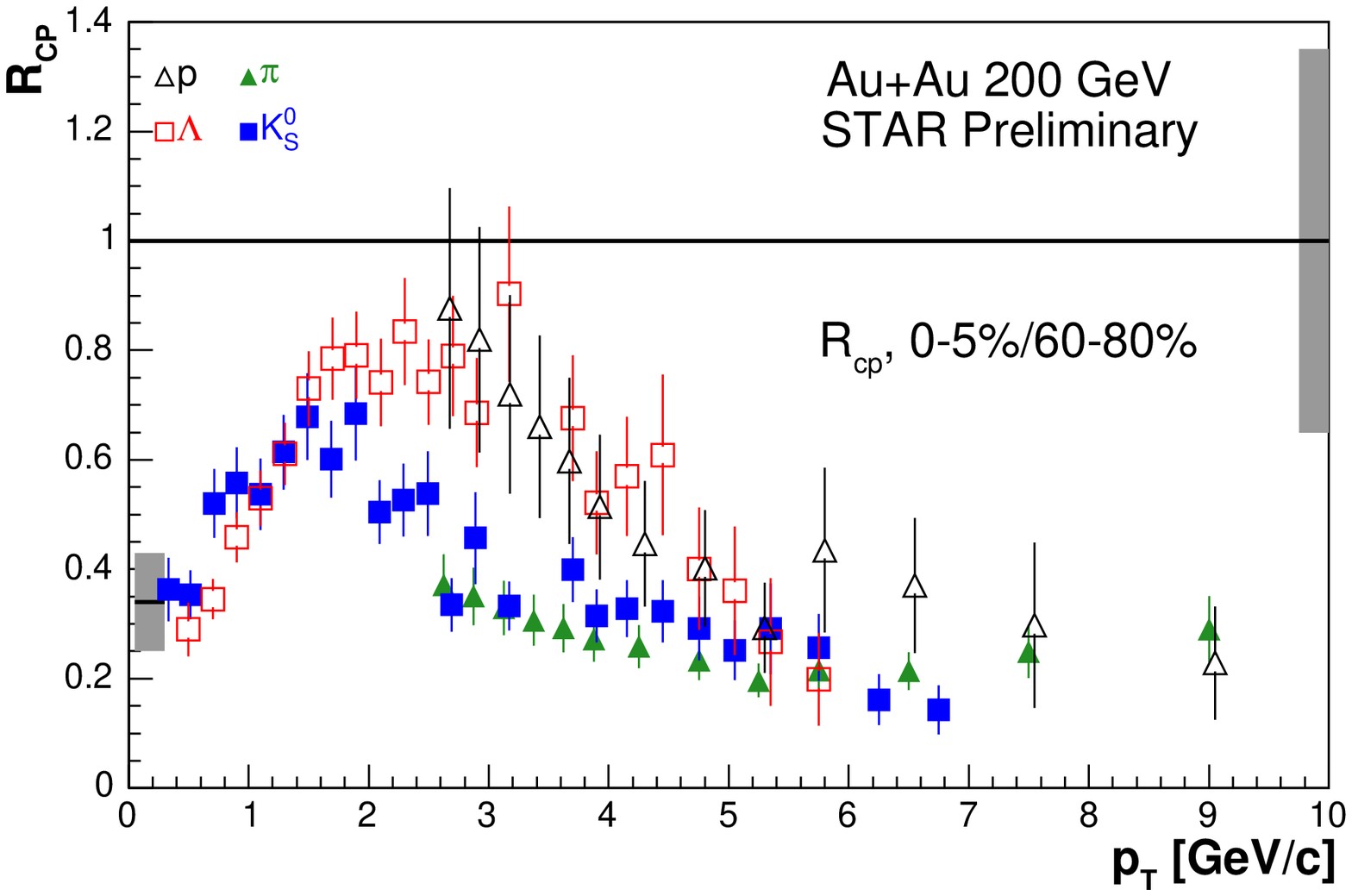}
 }
 \caption{(a) Proton to $\pi$ ratio as a function of $p_T$ for central 0-5\% Au+Au
collisions and p+p collisions.  For Au+Au, error bars are statistical and error bands
systematic uncertainties.  For p+p, error bars are combined statistical and systematic.
For further details, see ref.~\cite{barannikova}.(b) $R_{CP}$ as a function of $p_T$ for identified particles.
Errors are statistical and systematic, while the grey band at the right denotes common
scale uncertainties from $N_{binary}$.  See ref.~\protect\cite{barannikova} and \protect\cite{salur} for more details.} 
 \label{fig:ident}
\end{figure}

It has been known for some time now that 
at intermediate $p_T$ ($p_T<\sim$ 6 GeV/c) 
baryons behave differently in 
heavy ion collisions than mesons~\cite{Adams:2003am}.
This is shown clearly by figure~\ref{fig:protonpi}, which shows the 
ratio of proton to pion spectra in both central Au+Au collisions
and in p+p collisions.  The large enhancement in Au+Au collisions of the ratio
at intermediate $p_T$ indicates that the dominant source of particle production 
in this $p_T$ range is not jet fragmentation in vacuum.  With the high statistics
of run 4 and the full utilization of STAR's capabilities, this enhancement
is found to peak for $p_T\sim$ 2-3 GeV/c, beyond which the ratio falls
towards the ratio in p+p collisions, though with the current uncertainties
it is difficult to state conclusively if the ratio in Au+Au reaches that
seen in p+p.  

This enhancement also shows itself in the nuclear modification 
factor $R_{CP}$, the ratio between central and peripheral
Au+Au collisions of $N_{binary}$-scaled spectra.  
Results for $R_{cp}$ are shown in figure~\ref{fig:rcp}.
As with $v_2$~\cite{fqwang,mdoldenburg}, $R_{CP}$ separates into two groups,
baryons and mesons.  The $\phi$~\cite{caixz} and $K^*$~\cite{Adams:2004ep}, 
mesons with masses similar 
to the proton, follow the behavior of the mesons, proving that this separation
is not due to mass.  Such common 
grouping is violated in $R_{AA}$, in which the reference is 
from p+p collisions rather than peripheral collisions: strong enhancement
in strange baryon $R_{AA}$ is seen at intermediate $p_T$, with increasing
enhancement for increasing strangeness content~\cite{salur}.
There were indications in previous results that these 
dependences on hadron species 
disappeared at high $p_T$~\cite{Adams:2003am}, but the increased
reach of the run 4 dataset strengthens this conclusion, placing the application
of models incorporating parton energy loss on solid ground for $p_T>\sim$ 6 GeV/c.

\section{Heavy Flavor}
\begin{figure}
\centering
\includegraphics[width=0.47\textwidth]{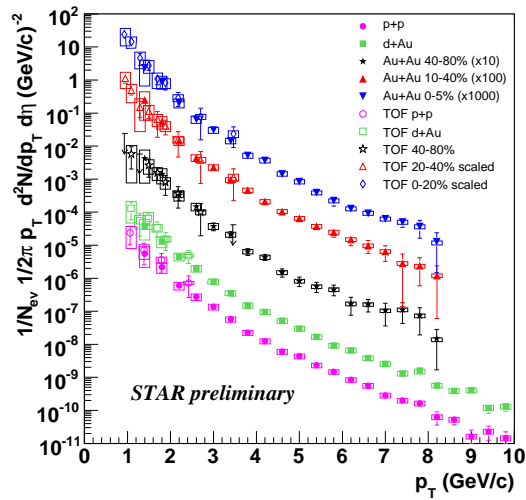}
\caption{Invariant differential yield vs. $p_T$ of non-photonic electrons.  Error bars are statistical, and boxes denote systematic uncertainties.  Spectra from 0-20\% and 20-40\% Au+Au collisions, measured by the TOF, have been scaled by $<N_{binary}>$ to match the centrality classes of the spectra from the BEMC (0-5\% and 10-40\%, respectively). See ref.~\protect\cite{zhang} and \protect\cite{bielcik} for more details.}
\label{fig:all_nonphotonic_spectra}
\end{figure}

The sector of charm and bottom hadrons is also accessible using the
identification capabilities of STAR.  At this conference,
we reported the first direct reconstruction
of D mesons in Au+Au collisions~\cite{zhang}.
Electrons can also be identified by using a combination of the
TPC dE/dx, the TOF patch, and the BEMC.  The dominant source of 
electrons in the detector is conversion of the photon
daughters of $\pi^0$ and $\eta$ hadrons in the detector material, 
along with Dalitz decays of these hadrons.  This ``photonic'' source can be
subtracted using an invariant mass technique.  The resulting ``non-photonic''
electrons are expected to be predominantly from the decay of charm
and bottom mesons.  Figure~\ref{fig:all_nonphotonic_spectra} shows the
resulting non-photonic electron spectra in p+p, d+Au, and Au+Au collisions.
The contribution to the electrons 
of charm relative to bottom is expected to decrease with increasing $p_T$, though without direct
measurement of the hadrons the point at which bottom dominates is somewhat
uncertain~\cite{Cacciari:2005rk}.  Since the radiative coupling of heavy quarks to the medium 
is expected to be smaller than that of light quarks, due to their mass, the
measurement of nuclear modification of heavy quarks is a sensitive test
of the picture of radiative energy loss.  

\begin{figure}
\centering
 \subfigure{
  \label{fig:raa_elec_tof}
  \includegraphics[width=0.47\textwidth]{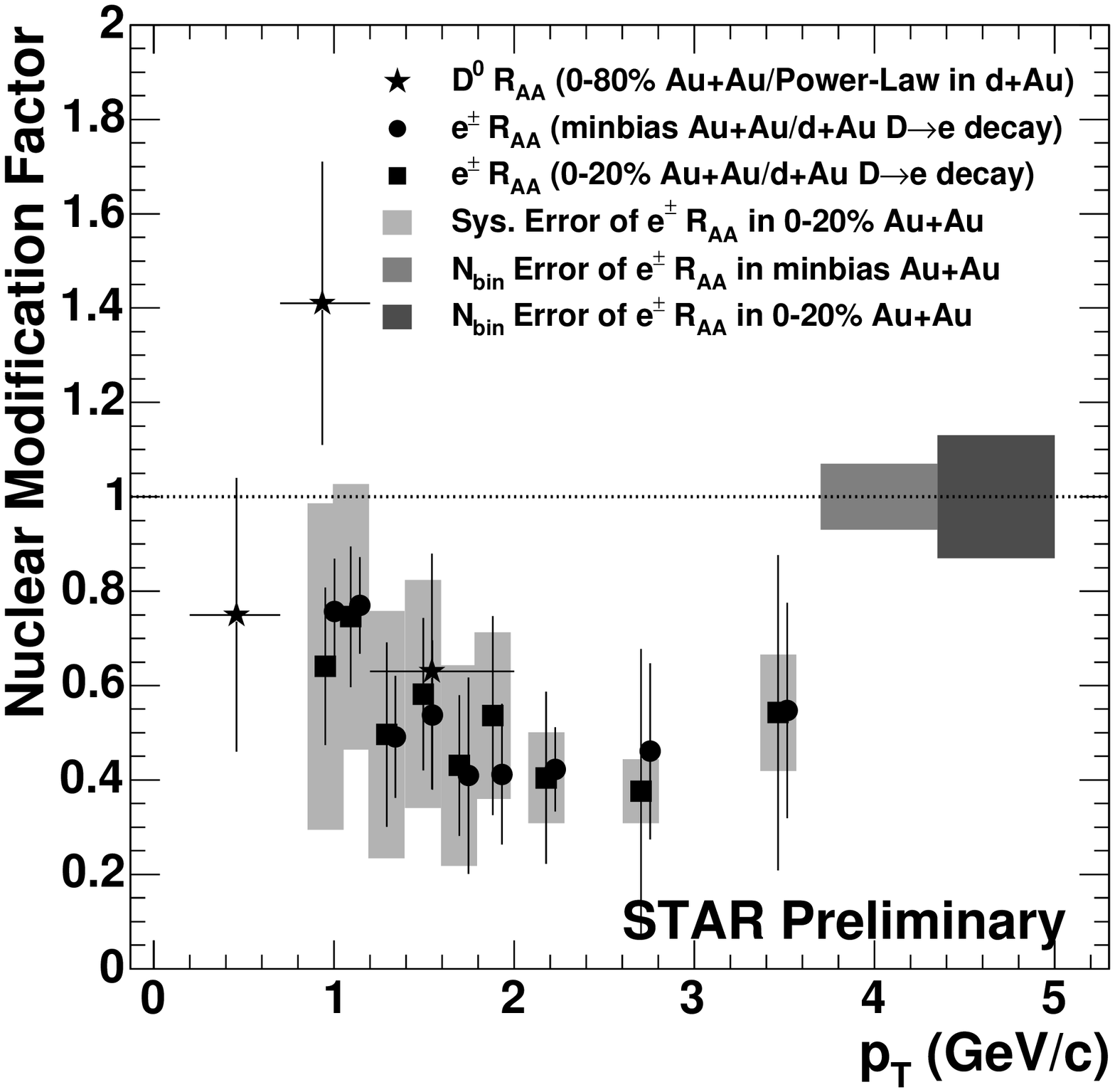}
  }
  \hfill
 \subfigure{
  \label{fig:raa_elec_bemc}
  \includegraphics[width=0.47\textwidth]{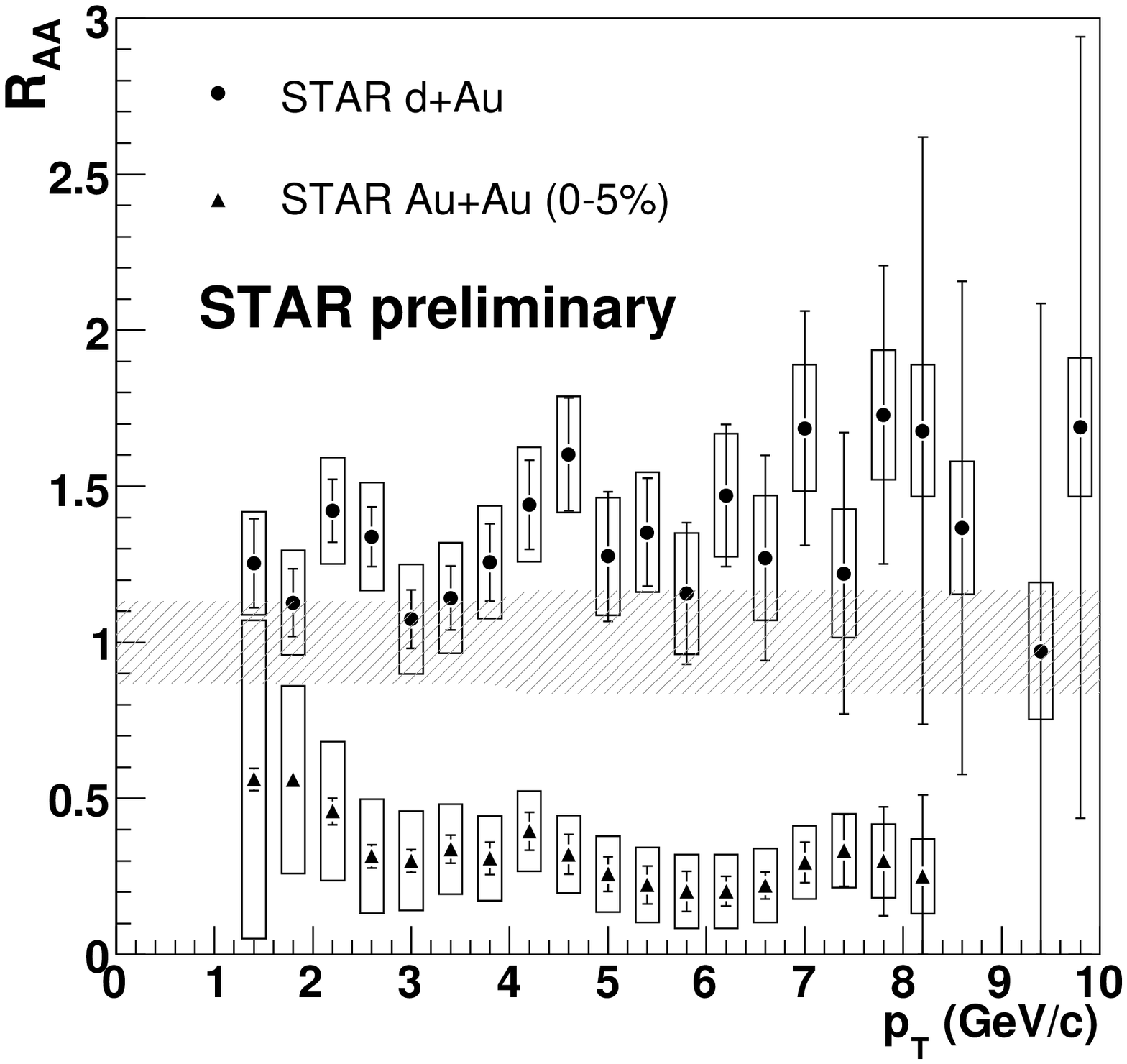}
  }

\caption{$R_{AB}$ as a function of $p_T$ for $D^0$ and non-photonic electrons. 
(a) $D^0$ and non-photonic electrons using the TOF.  See ref.~\protect\cite{zhang} 
for more details. (b) Non-photonic electrons using the BEMC.  Error bars
are statistical, error boxes point-to-point systematic uncertainties, and the band at unity denotes normalization uncertainty.  See ref.~\protect\cite{bielcik} for more details.}
\label{fig:raa_elec}
\end{figure}
The total cross-section of charm production, determined predominantly
by the direct measurement of D hadrons at low $p_T$, is found to scale
with $<N_{binary}>$, as expected for a hard probe produced in the
initial stages of the collision.  More differentially, 
figure~\ref{fig:raa_elec} shows the nuclear modification factor $R_{AA}$ 
for $D^0$ mesons and 
non-photonic electrons as a function of $p_T$ out to 8 GeV/c.  
For central Au+Au collisions, for $p_T>\sim$ 6 GeV/c,
$R_{AA}$ for non-photonic electrons is rather similar to  
that measured for charged hadrons (figure~\ref{fig:cucu_raa}).
This measurement stands in contrast to the predictions before the conference
of $R_{AA}\sim$0.5-0.6 in this $p_T$ range.
For extreme medium densities ($dN/dy_{gluon} = 3500$~\cite{Djordjevic:2005db}
or $\hat q = $ 14 GeV$^2$/fm~\cite{Armesto:2005iq}), this large level of
suppression can be reproduced within the radiative
framework, but only if the bottom contribution is assumed to be negligible.
Resolving whether this is a viable solution will depend on  
direct measurement of the charm and bottom contributions separately.   
Detector upgrades towards this goal are under active investigation~\cite{schweda}.

\begin{figure}
\centering
\includegraphics[width=0.7\textwidth]{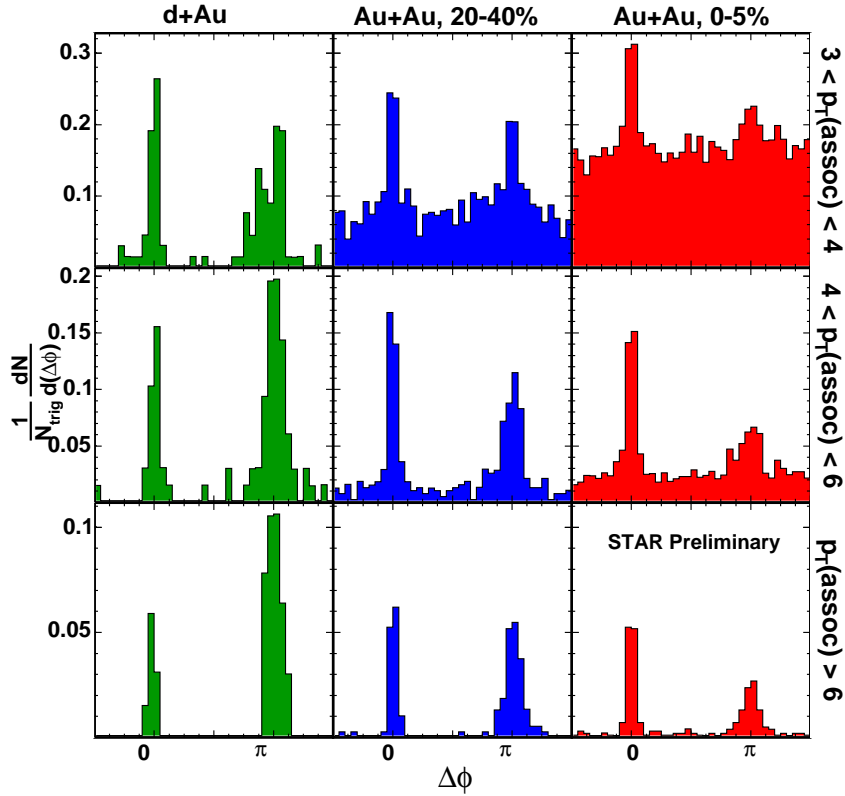}
\caption{Per-trigger azimuthal correlations of charged hadrons 
associated with a charged hadron trigger 
with $8<p_T<15$ GeV/c.  See ref.~\protect\cite{magestro} for more 
details.}
\label{fig:chargedhadron_corr}
\end{figure}

\begin{figure}
\centering
\includegraphics[width=0.5\textwidth]{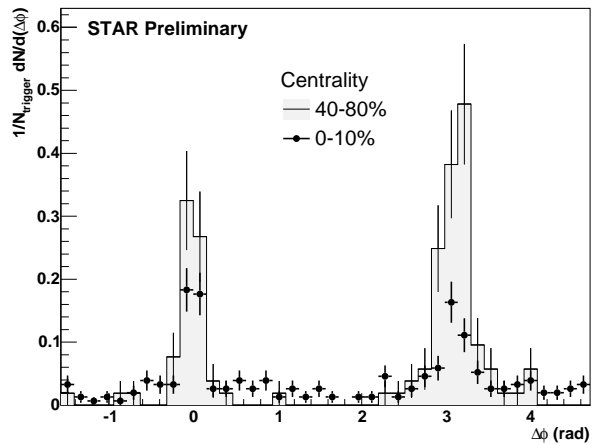}
\caption{Per-trigger azimuthal correlations of charged hadrons associated
with a trigger photon.  The trigger photon has $E_T>10$ GeV/c, while associated
charged hadrons have $4<p_T^{assoc}<E_T^{trigger}$.
See ref.~\protect\cite{dietel} for more details.}
\label{fig:photonhadron_corr}
\end{figure}

\begin{figure}[htb]
\centering
\includegraphics[width=0.8\textwidth]{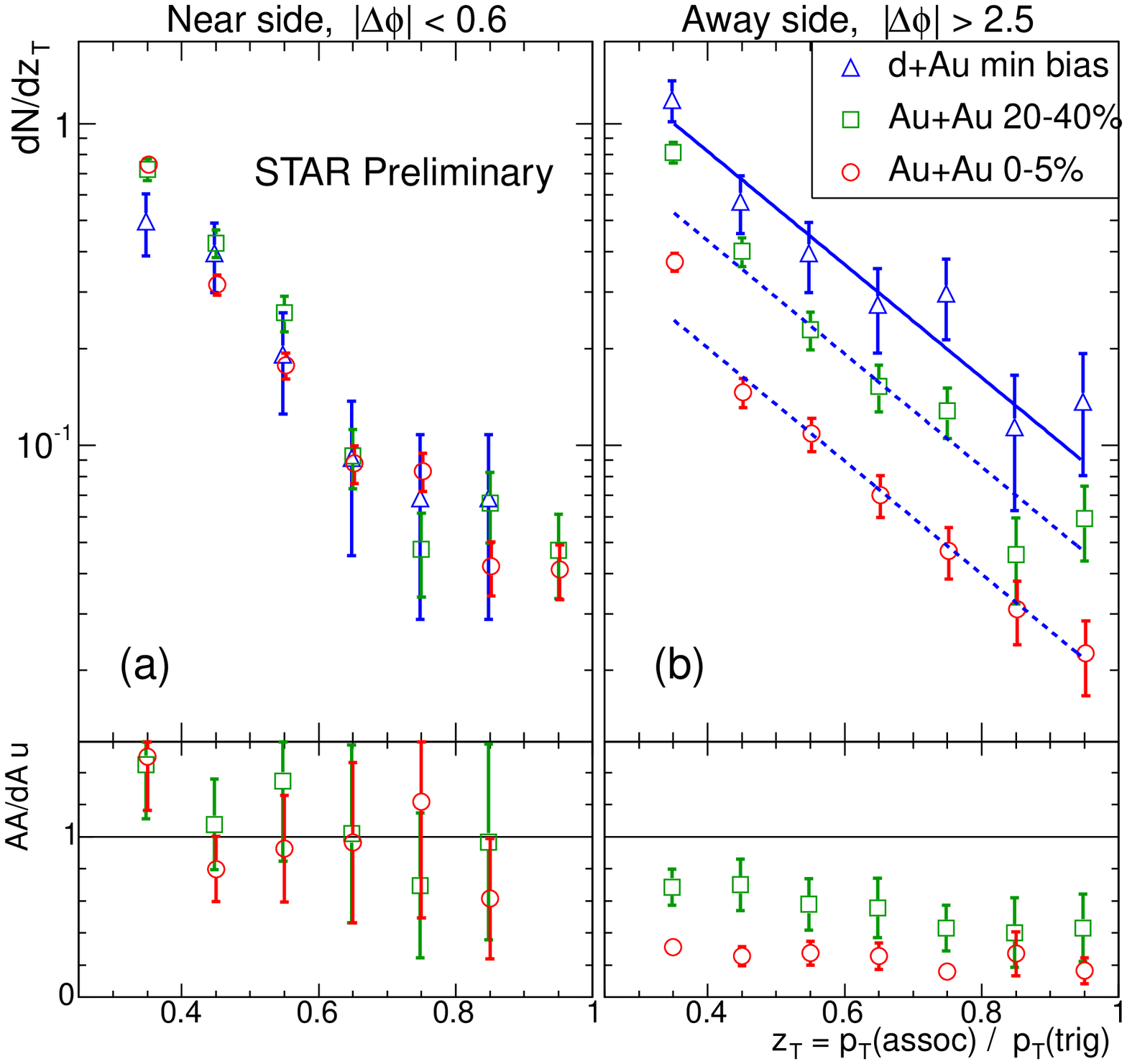}
\caption{Top: Dihadron fragmentation functions as a function of $z_T$, for 
(a) near-side and (b) away-side.  The solid line is an exponential fit
of the $z_T$ distribution for d+Au; the dashed lines are scaled by a factor
0.54(0.25) for 20-40\% (0-5\%) Au+Au.
Bottom: ratio of dihadron fragmentation functions
Au+Au/d+Au. See ref.~\protect\cite{magestro} for more details.}
\label{fig:iaa}
\end{figure}

\section{Dijet Correlations}
Dihadron correlations provide an alternative method to probe the medium than
single-particle spectra, with some advantages.  No Glauber calculation 
is required for the measurement of suppression,
and geometric and fragmentation biases are different than
those for single particles. 
However, previous analyses~\cite{Adler:2002tq,Adams:2005ph} 
have been limited by statistics to the intermediate
$p_T$ regime, where combinatoric backgrounds are high and multiple contributions
unrelated to jet fragmentation are present.  The increased statistics 
and experimental capabilities of the year 4 dataset have allowed STAR to increase the $p_T$ scale
and substantially reduce these issues.

Figure~\ref{fig:chargedhadron_corr}
shows per-trigger dihadron azimuthal correlations, in which both the
trigger and associated particle are charged hadrons and the yield
is normalized to the number of triggers.
The trigger hadron has $8<p_T<15$ GeV/c.
No background subtraction is performed in these plots.  
Both near-side (around $\Delta\phi=0$) and away-side (around $\Delta\phi=\pi$) 
peaks are visible, and for the highest associated $p_T$ combinatoric background
is negligible.  Little modification is apparent in the shape of the peaks,
though the height of the away-side peak shows a clear suppression 
in central Au+Au collisions relative to d+Au collisions.
These results stand in contrast to previous results in central
Au+Au collisions: for intermediate
thresholds, no significant strength was found in the away-side peak~\cite{Adler:2002tq},
while for extremely low thresholds strong modification in both the
shape and strength were observed~\cite{Adams:2005ph}.  
Unambiguously, these dihadron correlations reflect dijet phenomena
in central Au+Au collisions.

Figure~\ref{fig:photonhadron_corr} shows a related correlation, in which
the charged hadron trigger has been replaced by a photon detected in the
BEMC.  Here the trigger photon has $E_T>$ 10 GeV/c, and the associated
charged hadron has $4$ GeV/c $< p_T < E_T^{trigger}$.  For these
$E_T$, photons are expected to come both from fragmentation (through
the decay of $\pi^0$) and directly from the source~\cite{Filimonov:2005kp}.  In contrast to the  
charged-charged correlations, in the photon-charged correlation the strength 
of the near-side peak decreases from peripheral to central Au+Au collisions.
This may reflect a relative increase in the contribution of direct photons 
to the trigger particles in central collisions~\cite{dietel}.
This measurement represents a first step towards the ultimate goal
of the direct measurement of the medium modification of the
fragmentation function using direct photon-tagged hadrons.

With the clean correlation peaks of figure~\ref{fig:chargedhadron_corr},
one can form a dihadron ``fragmentation function'', as proposed in ref.~\cite{Wang:2003mm}.
Results are shown in figure~\ref{fig:iaa}.  The correlations are first binned
in $z_T = p_T^{assoc}/p_T^{trigger}$, and then the peaks on the near and away
side are integrated.  While the strength of the near-side peak shows little
modification from d+Au to central Au+Au collisions, the away-side peak shows a strong 
suppression, essentially independent of $z_T$ for large $z_T$.  Dihadrons
incorporate a completely different set of biases than single-particle spectra, yet
the level of suppression is similar to that seen for all single-particle spectra,
by a factor of four to five.  This similarity will place strong constraints
on the medium density inferred from energy-loss calculations, and may enable
the placement of an upper bound on the medium density.  It has been proposed that
from such an upper bound, combined with a lower bound on the entropy density,
a lower bound on the number of degrees of freedom of the medium can be obtained~\cite{Muller:2005en}.

\section{Conclusion}
Increased statistics and detector capability in run 4 have 
led to a qualitative improvement in the use of hard probes in STAR.
Light hadron as well as non-photonic electron spectra have been
measured to $p_T$ of up to 10 GeV/c: the surprising result is that
all hadrons, including those with heavy flavor, appear to be suppressed
to the same level in Au+Au collisions.  Dihadron correlations
have moved into the precision regime with clear dijet signatures,
and the beginning of the photon-jet program.  Interestingly,
dihadron correlations, which are affected by very different
biases than single particle spectra, show
the same level of suppression as single particle spectra.
With these results, and with more to come from the full analysis
of the current dataset and future datasets, it may be possible 
to move the determination of the properties of the medium
created at RHIC into the precision phase.


\begin{thebibliography}{9}
\bibitem{Adler:2002xw}
  C.~Adler {\it et al.}  [STAR Collaboration],
  %``Centrality dependence of high p(T) hadron suppression in Au + Au collisions
  %at s(NN)**(1/2) = 130-GeV,''
  Phys.\ Rev.\ Lett.\  {\bf 89} (2002) 202301.
%  [arXiv:nucl-ex/0206011].
  %%CITATION = NUCL-EX 0206011;%%

\bibitem{Adams:2003kv}
  J.~Adams {\it et al.}  [STAR Collaboration],
  %``Transverse momentum and collision energy dependence of high p(T) hadron
  %suppression in Au + Au collisions at ultrarelativistic energies,''
  Phys.\ Rev.\ Lett.\  {\bf 91} (2003) 172302.
  %[arXiv:nucl-ex/0305015].
  %%CITATION = NUCL-EX 0305015;%%

\bibitem{Adams:2003im}
  J.~Adams {\it et al.}  [STAR Collaboration],
  %``Evidence from d + Au measurements for final-state suppression of high  p(T)
  %hadrons in Au + Au collisions at RHIC,''
  Phys.\ Rev.\ Lett.\  {\bf 91} (2003) 072304.
  %[arXiv:nucl-ex/0306024].
  %%CITATION = NUCL-EX 0306024;%%

\bibitem{Adler:2002tq}
  C.~Adler {\it et al.}  [STAR Collaboration],
  %``Disappearance of back-to-back high p(T) hadron correlations in central Au +
  %Au collisions at s(NN)**(1/2) = 200-GeV,''
  Phys.\ Rev.\ Lett.\  {\bf 90} (2003) 082302.
%  [arXiv:nucl-ex/0210033].
  %%CITATION = NUCL-EX 0210033;%%


\bibitem{GLV} M. Gyulassy, P. Levai and I. Vitev, Nucl. Phys. A661 (1999) 637;
M. Gyulassy, P. Levai and I. Vitev, Nucl. Phys. B571 (2000) 197;
M. Gyulassy, P. Levai and I. Vitev, Phys. Rev. Lett. 85 (2000) 5535;
M. Gyulassy, P. Levai and I. Vitev, Nucl. Phys. B594 (2001) 371.

\bibitem{WW} E. Wang and X.N. Wang, Phys. Rev. Lett. 87 (2001) 142301.

\bibitem{BDMS} R. Baier, Y.L. Dokshitzer, A.H. Mueller, S. Peigne and D. Schiff, Nucl. Phys. B483 (1997) 291;
R. Baier, Y.L. Dokshitzer, A.H. Mueller and D.
Schiff, Nucl. Phys. B531 (1998) 403;
R. Baier, Y.L. Dokshitzer, A.H. Mueller and D. Schiff,  Phys. Rev. C60 (1999) 064902;
R. Baier, Y.L. Dokshitzer, A.H. Mueller and D. Schiff, JHEP 0109 (2001) 033.

\bibitem{Wied} U.A. Wiedemann, Nucl. Phys. B588 (2000) 303; U.A. Wiedemann, Nucl. Phys. A690 (2001) 731; C.A. Salgado and U.A. Wiedemann, Phys. Rev. Lett. 89 (2002) 092303;
C.A. Salgado, U.A. Wiedemann, Phys. Rev. D68 (2003) 014008.

%Fragile RAA
\bibitem{Dainese:2004te}
  A.~Dainese, C.~Loizides and G.~Paic,
  %``Leading-particle suppression in high energy nucleus nucleus collisions,''
  Eur.\ Phys.\ J.\ C {\bf 38} (2005) 461.
%  [arXiv:hep-ph/0406201].
  %%CITATION = HEP-PH 0406201;%%

\bibitem{Eskola:2004cr}
  K.~J.~Eskola, H.~Honkanen, C.~A.~Salgado and U.~A.~Wiedemann,
  %``The fragility of high-p(T) hadron spectra as a hard probe,''
  Nucl.\ Phys.\ A {\bf 747} (2005) 511.
%  [arXiv:hep-ph/0406319].
  %%CITATION = HEP-PH 0406319;%%

\bibitem{vanHees:2004gq}
  H.~van Hees and R.~Rapp,
  %``Thermalization of heavy quarks in the quark-gluon plasma,''
  Phys.\ Rev.\ C {\bf 71} (2005) 034907.
%  [arXiv:nucl-th/0412015].
  %%CITATION = NUCL-TH 0412015;%%
\bibitem{Moore:2004tg}
  G.~D.~Moore and D.~Teaney,
  %``How much do heavy quarks thermalize in a heavy ion collision?,''
  Phys.\ Rev.\ C {\bf 71} (2005) 064904.
%  [arXiv:hep-ph/0412346].
  %%CITATION = HEP-PH 0412346;%%
\bibitem{Mustafa:2004dr}
  M.~G.~Mustafa,
  %``Energy loss of charm quarks in the quark-gluon plasma: Collisional vs
  %radiative,''
  Phys.\ Rev.\ C {\bf 72} (2005) 014905.
%  [arXiv:hep-ph/0412402].
  %%CITATION = HEP-PH 0412402;%%

\bibitem{Adams:2005dq}
  J.~Adams {\it et al.}  [STAR Collaboration],
  %``Experimental and theoretical challenges in the search for the quark  gluon
  %plasma: The STAR collaboration's critical assessment of the  evidence from
  %RHIC collisions,''
  Nucl.\ Phys.\ A {\bf 757} (2005) 102.
  %[arXiv:nucl-ex/0501009].
 %%CITATION = NUCL-EX 0501009;%%

% \bibitem{Wang:1996pe}
%   X.~N.~Wang and Z.~Huang,
%   %``Medium-induced parton energy loss in gamma + jet events of  high-energy
%   %heavy-ion collisions,''
%   Phys.\ Rev.\ C {\bf 55}, 3047 (1997)
%   %[arxiv:hep-ph/9701227].
%   %%CITATION = HEP-PH 9701227;%%
\bibitem{Wang:1996yh}
  X.~N.~Wang, Z.~Huang and I.~Sarcevic,
  %``Jet quenching in the opposite direction of a tagged photon in  high-energy
  %heavy-ion collisions,''
  Phys.\ Rev.\ Lett.\  {\bf 77} (1996) 231.
%  [arXiv:hep-ph/9605213].
  %%CITATION = HEP-PH 9605213;%%

%Dead cone
\bibitem{Dokshitzer:2001zm}
  Y.~L.~Dokshitzer and D.~E.~Kharzeev,
  %``Heavy quark colorimetry of QCD matter,''
  Phys.\ Lett.\ B {\bf 519}, 199 (2001).
  %[arXiv:hep-ph/0106202].
  %%CITATION = HEP-PH 0106202;%%

\bibitem{Armesto:2005iq}
  N.~Armesto, A.~Dainese, C.~A.~Salgado and U.~A.~Wiedemann,
  %``Testing the color charge and mass dependence of parton energy loss with
  %heavy-to-light ratios at RHIC and LHC,''
  Phys.\ Rev.\ D {\bf 71} (2005) 054027; private communication.
  %[arXiv:hep-ph/0501225].
  %%CITATION = HEP-PH 0501225;%%

\bibitem{Djordjevic:2005db}
  M.~Djordjevic, M.~Gyulassy, R.~Vogt and S.~Wicks,
  %``Influence of bottom quark jet quenching on single electron tomography of Au
  %+ Au,''
  arXiv:nucl-th/0507019.
  %%CITATION = NUCL-TH 0507019;%%

\bibitem{star:nim} K.H. Ackermann et al., Nucl. Instrum. Methods A499 (2003) 624.

\bibitem{Adams:2004fc}
  J.~Adams {\it et al.}  [STAR Collaboration],
  %``Open charm yields in d + Au collisions at s(NN)**(1/2) = 200-GeV,''
  Phys.\ Rev.\ Lett.\  {\bf 94} (2005) 062301.
  %[arxiv:nucl-ex/0407006].
  %%CITATION = NUCL-EX 0407006;%%

\bibitem{Shao:2005iu}
  M.~Shao, O.~Barannikova, X.~Dong, Y.~Fisyak, L.~Ruan, P.~Sorensen and Z.~Xu,
  %``Extensive particle identification with TPC and TOF at the STAR
  %experiment,''
  arXiv:nucl-ex/0505026.
  %%CITATION = NUCL-EX 0505026;%%

\bibitem{Adams:2003yh}
  J.~Adams {\it et al.}  [STAR Collaboration],
  %``Production of charged pions and hadrons in Au + Au collisions at
  %s(NN)**(1/2) = 130-GeV,''
  arXiv:nucl-ex/0311017.
  %%CITATION = NUCL-EX 0311017;%%

\bibitem{Vitev:2002pf}
  I.~Vitev and M.~Gyulassy,
  %``High-p(T) tomography of d + Au and Au + Au at SPS, RHIC, and LHC,''
  Phys.\ Rev.\ Lett.\  {\bf 89} (2002) 252301; private communication.
  %[arXiv:hep-ph/0209161].
  %%CITATION = HEP-PH 0209161;%%


\bibitem{Adams:2003am}
  J.~Adams {\it et al.}  [STAR Collaboration],
  %``Particle dependence of azimuthal anisotropy and nuclear modification of
  %particle production at moderate p(T) in Au + Au collisions at  s(NN)**(1/2) =
  %200-GeV,''
  Phys.\ Rev.\ Lett.\  {\bf 92}, 052302 (2004).
%  [arXiv:nucl-ex/0306007].
  %%CITATION = NUCL-EX 0306007;%%

\bibitem{barannikova}
  O.~Barannikova {\it et al.} [STAR Collaboration], these proceedings.

\bibitem{salur}
  S.~Salur {\it et al.} [STAR Collaboration], these proceedings.

\bibitem{fqwang}
  F.~Wang {\it et al.} [STAR Collaboration], these proceedings.

\bibitem{mdoldenburg}
  M.~Oldenburg {\it et al.} [STAR Collaboration], these proceedings.

\bibitem{caixz}
  X.Z.~Cai {\it et al.} [STAR Collaboration], these proceedings.

\bibitem{Adams:2004ep}
  J.~Adams {\it et al.}  [STAR Collaboration],
  %``K(892)* Resonance Production in Au+Au and p+p Collisions at sqrt(sNN) = 200
  %GeV at STAR,''
  Phys.\ Rev.\ C {\bf 71} (2005) 064902.
%  [arXiv:nucl-ex/0412019].
  %%CITATION = NUCL-EX 0412019;%%

\bibitem{zhang}
H.~Zhang {\it et al.} [STAR Collaboration], these proceedings.

\bibitem{Cacciari:2005rk}
  M.~Cacciari, P.~Nason and R.~Vogt,
  %``QCD predictions for charm and bottom production at RHIC,''
  arXiv:hep-ph/0502203; R.~Vogt, these proceedings. 
  %%CITATION = HEP-PH 0502203;%%

\bibitem{bielcik}
J.~Bielcik {\it et al.} [STAR Collaboration], these proceedings.

\bibitem{schweda}
K.~Schweda {\it et al.} [STAR Collaboration], these proceedings.

\bibitem{Lin:2003jy}
  Z.~w.~Lin and D.~Molnar,
  %``Quark coalescence and elliptic flow of charm hadrons,''
  Phys.\ Rev.\ C {\bf 68} (2003) 044901.
%  [arXiv:nucl-th/0304045].
  %%CITATION = NUCL-TH 0304045;%%

\bibitem{Greco:2003vf}
  V.~Greco, C.~M.~Ko and R.~Rapp,
  %``Quark coalescence for charmed mesons in ultrarelativistic heavy-ion
  %collisions,''
  Phys.\ Lett.\ B {\bf 595} (2004) 202.
%  [arXiv:nucl-th/0312100].
  %%CITATION = NUCL-TH 0312100;%%

\bibitem{Adams:2005ph}
  J.~Adams {\it et al.}  [STAR Collaboration],
  %``Distributions of charged hadrons associated with high transverse  momentum
  %particles in p p and Au + Au collisions at s(NN)**(1/2) =  200-GeV,''
  Phys.\ Rev.\ Lett.\  {\bf 95} (2005) 152301.
%  [arXiv:nucl-ex/0501016].
  %%CITATION = NUCL-EX 0501016;%%

%\cite{Filimonov:2005kp}
\bibitem{Filimonov:2005kp}
 K.~Filimonov,
 %``High p(T) correlations of gamma and charged hadrons at RHIC,''
 arXiv:nucl-ex/0505008.
 %%CITATION = NUCL-EX 0505008;%% 

\bibitem{dietel}
T.~Dietel {\it et al.} [STAR Collaboration], these proceedings.

\bibitem{magestro}
D.~Magestro {\it et al.} [STAR Collaboration], these proceedings.

\bibitem{Wang:2003mm}
  X.~N.~Wang,
  %``High p(T) hadron spectra, azimuthal anisotropy and back-to-back
  %correlations in high-energy heavy-ion collisions,''
  Phys.\ Lett.\ B {\bf 595} (2004) 165.
 %[arXiv:nucl-th/0305010].
  %%CITATION = NUCL-TH 0305010;%%

\bibitem{Muller:2005en}
  B.~Muller and K.~Rajagopal,
  %``From entropy and jet quenching to deconfinement?,''
  arXiv:hep-ph/0502174.
  %%CITATION = HEP-PH 0502174;%%




\end{thebibliography}
\end{document}